C. Malesios[1]

[1]Department of Rural Development, Democritus University of Thrace, 193 Pantazidou Str., Orestiada, Greece, e-mail: malesios@agro.duth.gr

# Measuring the robustness of the journal h-index with respect to publication and citation values: A Bayesian sensitivity analysis

**Abstract**

*Braun et al. (2006)* recommended using the *h*-index as an alternative to the journal impact factor (IF) to qualify journals. In this paper, a Bayesian-based sensitivity analysis is performed with the aid of mathematical models to examine the behavior of the journal *h*-index to changes in the publication/citation counts of journals. Sensitivity of the *h*-index was most apparent for changes in the number of citations, revealing similar patterns of behavior for almost all models and independently to the field of research. In general, the *h*-index was found to be robust to changes in citations up to approximately the $25^{th}$ percentile of the citation distribution, inflating its value afterwards.

## 1 Introduction

*Hirsch (2005)* introduced the *h*-index for the assessment of the research performance of scientists. Not only the indicator has found a wide use in a very short time, but also a series of articles were subsequently published either proposing modifications of the original *h*-index for its improvement, or new implementations of the proposed index. Increasingly, the *h*-index is proposed as an alternative to the most commonly used IF for evaluating the scientific impact of journals (see, e.g. *Bornmann et al., 2012*; *Malesios and Arabatzis, 2012*; *Schubert, 2015*). Despite the fact that various mathematical models for the *h*-index have been proposed, yet little is known about the mechanisms governing the relationship between the *h*-index and publications (*P*)/citations (*C*) and its robustness to the latter indicators by utilizing the aforementioned models. The general perspective is that the (journal) *h*-index is robust to changes in number of publications and citations. *Franceschini et al. (2013)* for instance deduce that *h*-indices are robust to small variations in the publication/citation data and even to significant changes in the *C* values of the papers of interest, by investigating the robustness of the *h*-index to missing or wrong citation records. *Courtault and Hayek (2008)* have theoretically shown that a significant number of papers significantly cited must be published to increase the *h*-index. In the same lines, *Rousseau (2007)* found, by utilizing theoretical models, that a relative small number of highly cited publications have a small influence on the *h*-index. According to *Minasny et al. (2013)*, the *h*-index is less sensitive to the increase in the number of citations and it does not penalize a journal for publishing a larger number of papers. For a more applied examination concerning the robustness of the *h*-index we refer the interested reader to *Vanclay (2007)*.

However, the latter claims have not been examined thoroughly up to now, especially in the context of the *h*-index of a research journal. One may ask: what are small variations in *P*, *C* and how they can be quantified? This paper tries to fill this gap and answer the following question; how the *h*-index varies according to specific changes in the number of *P* and *C*? This research question cannot be addressed without specifying a mathematical relation between *h*, *P* and *C*. Hence, by relying on some of the well-established mathematical functions relating *h*-index with *P*, *C,* an empirical contribution to the issue of quantifying the sensitivity of the *h*-index by adopting a statistical modeling view is attempted, within the Bayesian paradigm. Bayesian methods permit model flexibility and appropriateness and the present study shall attempt to highlight the practical benefits of the Bayesian view of statistics.

In this context, it shall be also attempted to answer which model is more robust when compared to the others. The proposed methodology is illustrated utilizing two different datasets consisting of the *h*-indices, *P* and *C* of the journals in the fields of ecology and forestry included in the Web of Science (WoS) (Collection date: March, 2013 and November 2011 for ecology and forestry journals respectively). The total samples consisted of 264,519 and 71,683 research publications from 134 ecology and 54 forestry, scientific journals, respectively, thus constituting two diverse groups of data suitable for credible inferences. For more details on the collected data see *Malesios (2015)*.

**2 Methods**

*2.1 Introduction to Bayesian model-based inference*

Statistics uses two major paradigms, classical (or conventional or frequentist) and Bayesian. Bayesian methods can incorporate scientific hypothesis in the analysis through the prior distribution and also have the advantage of being applied to problems with too complex structures that cannot be solved through classical statistics (*Bernardo, 2003*). Many statistical models are currently too complex to be fitted using classical statistical methods, but they can be fitted using Bayesian computational methods.

Inference for classical statistical modelling traditionally is based on the Maximum Likelihood (ML), where parameter estimates and corresponding confidence intervals are valid only for large samples. In contrast, Bayesian inference is exact for any sample, regardless of its size. Another distinctive characteristic of the Bayesian paradigm is that the data are treated as a fixed quantity and the parameters as random variables. Hence, in this sense, every parameter is assigned distributions, in contrast to classical statistics, where parameters are treated as fixed unknown constants.

Although Bayesian inference has been criticized for the use of the prior distribution, alternatively to utilizing an informative prior distribution it is also possible to specify ignorance in Bayesian analysis (i.e. we do not know anything about the parameters of interest) by assigning an uninformative (or vague or diffuse) prior. By the term uninformative prior, we mean assigning to the parameter a prior distribution with a very large variance.

In a general setting, under Bayesian inference we denote by $\boldsymbol{\theta} = (\theta_1, \theta_2, ..., \theta_k)^t$ the vector of a set of, say *k*, unobserved parameters, and by $\mathbf{x}$ the observed data. Bayesian inference is based on Bayes' theorem, according to which the posterior distribution, denoted by $p(\boldsymbol{\theta} \mid \mathbf{x})$ is given by:

$$p(\boldsymbol{\theta} \mid \mathbf{x}) = \frac{p(\mathbf{x} \mid \boldsymbol{\theta}) \cdot p(\boldsymbol{\theta})}{p(\mathbf{x})} = \frac{p(\mathbf{x} \mid \boldsymbol{\theta}) \cdot p(\boldsymbol{\theta})}{\int_{\boldsymbol{\theta}} p(\mathbf{x} \mid \boldsymbol{\theta}) \cdot p(\boldsymbol{\theta}) d\boldsymbol{\theta}}. \tag{1}$$

Equation (1) states that the probability of parameters $\boldsymbol{\theta}$ given the data $\mathbf{x}$ is proportional to the likelihood function $L(\boldsymbol{\theta}) = p(\mathbf{x} \mid \boldsymbol{\theta})$ and the prior distribution of $\boldsymbol{\theta}$, $p(\boldsymbol{\theta})$, i.e.:

posterior $\propto$ likelihood $\times$ prior.

The latter constitutes the intuitive basis of model-based Bayesian inference combining the information that we know before (through prior distribution), updated using the likelihood function (the data) in order to obtain the posterior distribution which gives information about the parameter of interest.

The most challenging issue in Bayesian inference is - as is well known – the normalizing term $p(\mathbf{x})$ (often called the marginal likelihood) in the denominator of equation (1), due to that in most modelling cases $p(\mathbf{x})$ includes complex high-dimensional integrals which are analytically intractable. Due to this issue, the problem of generating samples from the posterior distribution $p(\boldsymbol{\theta} \mid \mathbf{x})$ is not straightforward. Only after the mid-1980s the implementation of simulation-based computing algorithms like Markov chain Monte Carlo (McMC) (*Gelman et al., 2003*) on widely accessible powerful computers helped to overcome these problems and led to an explosion of interest in Bayesian modelling (*Ntzoufras, 2011*).

Markov chain simulation yields a sample from the posterior distribution $p(\boldsymbol{\theta} \mid \mathbf{x})$ of a parameter. One of the most widely used McMC techniques is Gibbs sampler (*Geman and Geman, 1984; Gelfand et al., 1990*). A brief description of the Gibbs sampler iterative scheme for obtaining posterior samples for parameters $\boldsymbol{\theta}$ is presented below:

| |
|---|
| **(a)** Choose initial values for $\boldsymbol{\theta} = (\theta_1, \theta_2, ..., \theta_k)^t$ : $\theta_1^{(0)}, \theta_2^{(0)}, ..., \theta_k^{(0)}$ |
| **(b)** Sample value $\theta_1^{(1)}$ from the conditional distribution $p(\theta_1 \mid \theta_2^{(0)}, \theta_3^{(0)}, ..., \theta_k^{(0)}, \mathbf{x})$, $\theta_2^{(1)}$ from $p(\theta_2 \mid \theta_1^{(1)}, \theta_3^{(0)}, ..., \theta_k^{(0)}, \mathbf{x})$, ...., $\theta_k^{(1)}$ from $p(\theta_k \mid \theta_1^{(1)}, \theta_2^{(1)}, ..., \theta_{k-1}^{(1)}, \mathbf{x})$. |

> **(c)** Repeat step (b) an adequate number of iterations (e.g. N=10000, 100000) in order to obtain the posterior distribution of each $\theta_i$ $(i = 1, 2, ..., k)$.

During these iterative schemes, the monitoring of the convergence of the algorithm is required. From a theoretical point of view, convergence of the algorithm implies convergence to the stationary distribution of the parameter, which in this case of Bayesian inference is the posterior distribution. In practice, there are several diagnostics for checking this, both visual and more formal. Trace plots (or history plots) are for example one visual way for examining convergence. If the chains of posterior samples, plotted against the sampling time, appear stable in certain areas of the parameter space, there is indication of convergence. For a detailed presentation of formal diagnostic techniques for assessing convergence we refer the interested reader to *Brooks and Roberts (1998)*.

When convergence is reached then after discarding the first $B$ posterior samples we then consider the remaining $\{\boldsymbol{\theta}^{(B+1)}, \boldsymbol{\theta}^{(B+2)}, ..., \boldsymbol{\theta}^{(N)}\}$ samples. Summary measures, such as the median or the mean of the posterior distribution can be used as point estimates for the parameters $\boldsymbol{\theta}$, whereas the $\alpha/2$ and $1-\alpha/2$ posterior quantiles can be used to construct the $(1-\alpha)100\%$ posterior credible intervals for the $\boldsymbol{\theta}$s.

*2.2 Bayesian methods in scientometrics*

In the field of scientometrics in general – and particularly of research assessment with *h*-type indices – techniques from Bayesian statistics have hardly been used. Hence, the utilization of a Bayesian statistical modelling view may also serve as a guideline for future research applying Bayesian inference.

The few examples of applying Bayesian methods in scientometrics include the paper of *Ibáñez et al. (2011)*, utilizing Bayesian networks to examine relationships between bibliometric indices (see also the recent publication of *Bornmann et al. (2016)* presenting applications based on Bayesian networks) and *Malesios (2015)* performing parameter estimation from a Bayesian modeling perspective, by statistically fitting the most well-established mathematical models for the *h*-index using a McMC sampling scheme. The proposed by *Malesios (2015)* methodology was illustrated utilizing the two previously presented journal bibliometric data and assuming typical alternative distributions for count data such as the values of the *h*-index. The results revealed significant differences among the fitted models as concerns their fit. Most suitable models for the association between the *h*-index and *P*, *C* were found to be the Glänzel-Schubert model (*Schubert and Glänzel, 2007*) and a three-parameter Hirsch model[1] *(Ye, 2011)*, whereas the negative binomial (NB) distribution assumed for the *h*-index values has been found to be a useful alternative to the commonly utilized Gaussian distribution.

For the current analysis the best fitted models as were identified in *Malesios (2015)* are utilized, i.e. the Gaussian Glänzel-Schubert model (*Schubert and Glänzel, 2007*), and the three-parameter Hirsch model *(Ye, 2011)* assuming a Gaussian and a NB distribution for the *h*-index values.

In addition, based on further research conducted by the author, an additional mathematical function is included that was found to perform similarly with the three previously described models. The proposed function is a modification of the classic Lotkaian model $h = P^{1/\alpha}$ of *Egghe and Rousseau (2006)* (see also *Franceschini et al., 2013*) that depends on *C* and a single parameter $\alpha$ through:

$$h = \left(\frac{\alpha-2}{\alpha-1}A\right)^{1/\alpha}, \qquad (2)$$

where $A = \dfrac{C}{\alpha-2} = P\dfrac{\alpha-1}{\alpha-2}$ and $\alpha > 2$ (see Table 1 for the mathematical forms of the fitted statistical models in the current analysis). Attempts to fit models based on a dynamic temporal *h*-index (see *Egghe, 2009*) resulted in poor fit and thus the latter models were excluded from further analyses.

Specifically, the theoretical mathematical model functions of Table 1 are utilized in order to fit Bayesian regression-type models of the following general form:

$$\begin{aligned}
H_i &\sim f(h_i \mid \theta_i) \\
\theta_i &= h(\mu_i) \\
\mu_i &= g(P,C)
\end{aligned} \qquad (3)$$

---

[1] In *Malesios (2015)* a slightly varied version of the three-parameter Hirsch model used here has been utilized, assuming $\alpha = a$, however the results of both analyses were very similar with no significant variations.

where $H_i$ is the random variable of the theoretical *h*-index following distribution $f$ (i.e. either the Gaussian or the Negative Binomial distribution), $\theta_i = h(\cdot)$ denotes the link function of the mean *h*-index to each of the functions in Table 1 (being $h(\mu_i) = \mu_i$ and $h(\mu_i) = \frac{r(1-q_i)}{q_i}$ for the Gaussian and the negative binomial distribution, respectively), and finally $g(\cdot)$ denotes each of the 4 theoretical functions tested. ($i = (1,2,...,130)$ and $i = (1,2,...,54)$ for the ecology and forestry journals, respectively).

(TABLE 1 APPROXIMATELY HERE)

As already described in Section 2.1, in order to implement Bayesian inference concerning the parameters of the four *h*-index regression-type models, one has to specify both the likelihood and the prior distributions for the models' parameters. Then – after obtaining the posterior distribution – summary measures and density plots can be used to describe the posterior parameter estimates. To illustrate this, expressions (4) to (7) describe the Glänzel-Schubert Gaussian model in Bayesian notation:

$$h_i \sim Normal\left(cP_i^{1/\alpha+1} \cdot \left(\frac{C_i}{P_i}\right)^{\alpha/\alpha+1}, \sigma\right), \tag{4}$$

priors:

$$\alpha \sim Normal(1, 100), \tag{5}$$

$$c \sim Normal(0, 100), \tag{6}$$

$$\tau \sim Gamma(0.001, 0.001). \tag{7}$$

where $\tau = 1/\sigma$ the precision and subscript *i* the sample size. Hence, the parameters $\alpha$ and $c$ are given independent non-informative priors with mean one and zero respectively and with some precision. Additionally, the precision is picked from a Gamma distribution with a specified mean and shape parameter.

Figure A1 in the Appendix shows part of the WinBUGS program (*Lunn et al., 2000*) that describes the prior distribution of the parameters and the likelihood of the data for one of the fitted models (the Glänzel-Schubert model – ecology field).

*2.3 Sensitivity Analysis*

Sensitivity analysis is an extremely useful methodological tool (*Saltelli et al., 2004*). A number of studies have previously demonstrated that "goodness of fit" alone is insufficient in reliably classifying the credibility of a model and sensitivity analysis is commonly employed as a secondary method for evaluating the suitability of a particular model (*Saithong et al., 2010*). In this stage we proceed with exploring the effect of *P* and *C* on the *h*-index via suitable sensitivity analysis techniques. Inference with the proposed scheme is implemented by McMC posterior simulation, allowing formal sensitivity analysis with respect to prior and likelihood assumptions of the original model analysis. Specifically, inference for alternative scenarios is derived via the implementation of a multivariate probabilistic sensitivity analysis-type scheme, where sensitivity analysis is obtained by increasing each parameter by a given percentage while leaving all others constant and quantifying the change in the model's output. By this, it is made possible to quantify in an accurate way the magnitude of the effects each one of *P* and *C* has on the *h*-index. In this context, we utilize distinct values for either *P* or *C* keeping the other covariate fixed at its median value for the model including both *P* and *C*, and then sample from the posterior density of the response by sampling from the posterior of the covariates' coefficient estimates. The use of the posterior samples obtained by the Bayesian inference instead of fixed estimates for the parameters of the mathematical models constitutes the probabilistic nature of the proposed methodology. There are certain advantages when following this approach; the probabilistic sensitivity analysis under a Bayesian framework proposed in this paper is chosen for certain desirable properties such as the ability to introduce uncertainty inherent in the parameters of interest, by using samples from their posterior distributions instead of the standard static sensitivity analysis where arbitrary values are usually assigned to them. This feature is desirable at least for the mathematical models examined here since that some of the parameters are not unknown but restricted to certain intervals. Furthermore, when multiple parameters are estimated from the same model, it is likely that correlations will be induced between them. Probabilistic sensitivity analyses schemes adequately account for the latter since that by using the posterior output we preserve the potential correlation structure, an important aspect especially in estimating (non)-linear functionals. To implement this scheme we use the posterior samples for the parameters derived from the fit of the Bayesian models based on the functions of Table 1. Hence, we conduct Bayesian posterior estimation and then export the posterior samples of parameters of interest to another statistical program that allows implementing the probabilistic sensitivity analysis.

Sensitivity of the *h*-index for our analysis is visually determined by plots of input vs. the output values and more formally by utilizing the sensitivity index (*SI*) given by:

$$SI = \frac{h_{\max} - h_{\min}}{h_{\max}}, \tag{4}$$

where $h_{\max}$ and $h_{\min}$ denote the maximum and minimum output values resulting from varying the input over its entire range (*Hoffman and Gardner, 1983*). *SI* ranges between 0 and 1, with larger values indicating higher sensitivity.

*2.4 Inference*

Bayesian inference is implemented with the use of an McMC sampling algorithm, as described previously. For specifying the priors, low information prior distributions were used (i.e. Gaussian distributions with zero mean and very large variance), truncated though accordingly to comply with the theoretical specification range for each parameter as is shown in Table 1. To fit the theoretical functions under the Bayesian framework the WinBUGS software was used. WinBUGS uses McMC methods to generate samples from the posterior distribution of the specified models.

Model selection was performed based upon Bayesian model comparison and model selection criteria. Specifically, the mean of the posterior deviance ($\bar{D}$) is utilized, based upon the deviance:

$$D(\theta) = -2\log p(x\mid\theta), \tag{5}$$

where $p(x|\theta)$ is the likelihood and $\theta$ some parameter. According to equation (5), high values of $D(\theta)$ indicate low values for the likelihood which accordingly means that the model does not fit the data well. From a Bayesian perspective, since the deviance $D(\theta)$ is a function of the $\theta$ parameter which is assumed as having a posterior distribution, $D(\theta)$ also has a posterior distribution. Averaging the posterior deviance values obtained by McMC we obtain the mean of the posterior deviance, with smaller values being preferable (Full details on $\bar{D}$ can be found in *Spiegelhalter et al., 2002*).

The posterior results for the models' parameters have been obtained by using 5,000 iterations as initial burn-in period and an additional sample of 50,000 iterations. Sensitivity analyses are performed with the use of the R software (*R Core Team, 2012*). The WinBUGS code for all fitted models and the data as well as the R code for performing sensitivity analyses is available upon request by the author.

**3 Results and Discussion**

Table 2 provides the fit statistics for the analyzed models under Bayesian inference. As is observed, the Egghe-Rousseau model under a negative binomial distribution is a competitive alternative to the other three best fitted models identified in previous analyses (*Malesios, 2015*), especially for the forestry data. However best fit is shown by the Glänzel-Schubert model for both journal categories, as shown by the mean deviance values $\bar{D}$.

(TABLE 2 APPROXIMATELY HERE)

The corresponding posterior distributions of the deviance values for the eight fitted models are presented in the following violin plots (Figure 1). The plots provide an additional visual indication as concerns the selection of the best model.

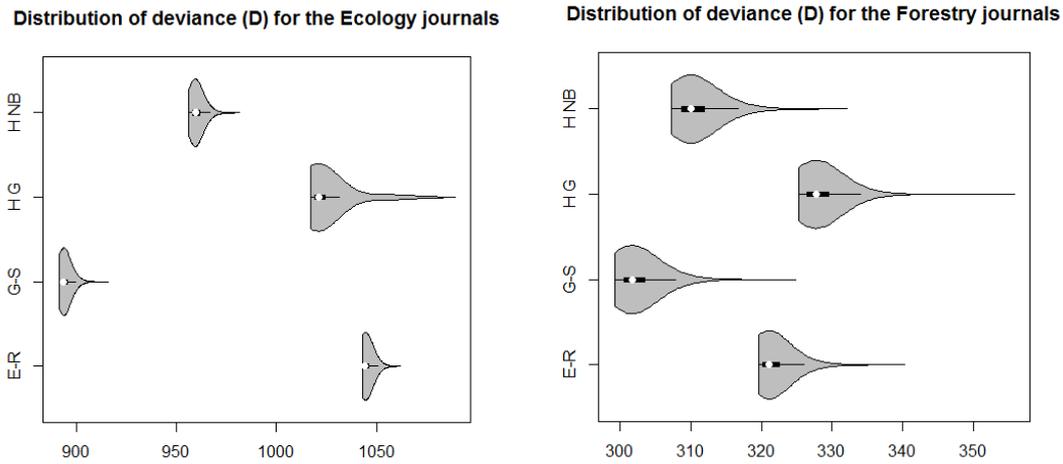

**Fig 1.** *Violin plots of the distribution of deviance values for the fitted models*

Posterior medians for the models' parameters are summarized in Table 3.

(TABLE 3 APPROXIMATELY HERE)

History plots of the posterior samples of parameters (see for example Figure 2 for the parameters $\alpha$ and $c$ of the Glänzel-Schubert model) indicated no lack of convergence for all fitted models.

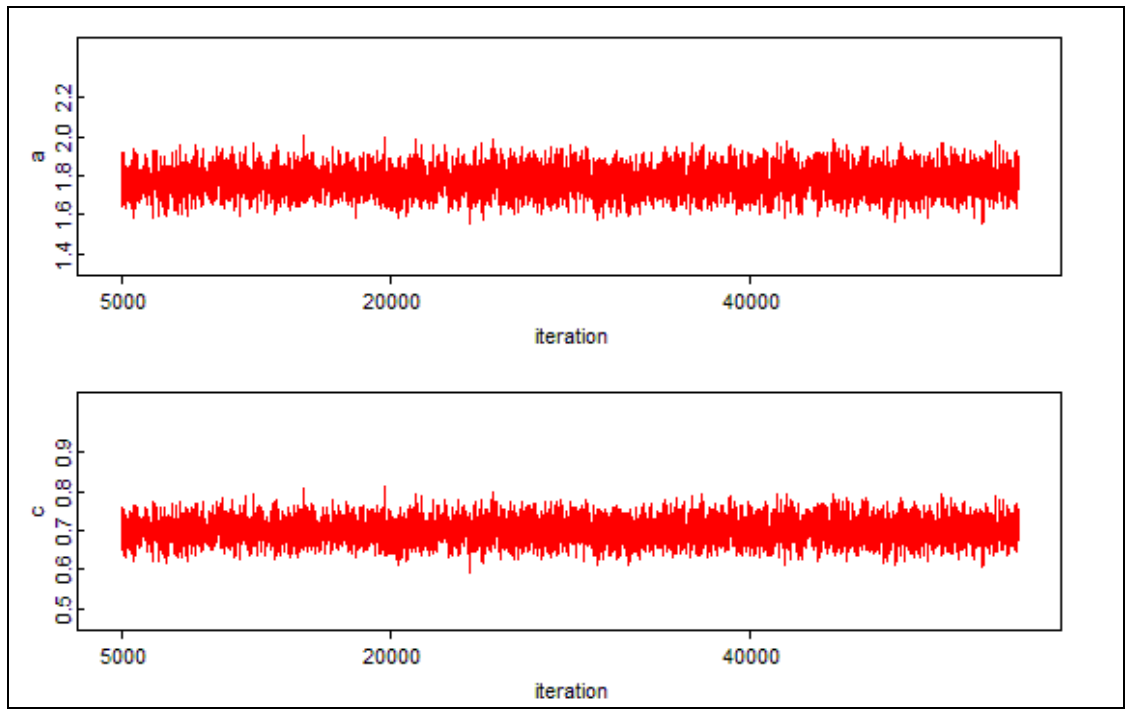

**Fig 2.** *History plots for the posterior parameter estimates of the best fitted model (Glänzel-Schubert model – ecology field) estimated from 50,000 iterations.*

Next we proceed with the sensitivity analysis results as described in section 2.1, performed for the four selected theoretical models. Two separate sensitivity analyses are performed, first by utilizing the complete data output on *P*, *C* (see Tables A1 and A2 in the Appendix for the descriptive statistics of the three variables' values used for the analysis) for the two journal categories, and subsequently using artificial values of the two variables to examine locally the sensitivity of *h* using distinct selected values for *P*, *C* near their medians (i.e. the range between -30% and 30% of *P*, *C* values around their medians with increments of 5%). Tables 4 and 5 show the values of *SI*s for the complete publication/citation data and the - local to the median - values, respectively.

(TABLE 4 APPROXIMATELY HERE)

(TABLE 5 APPROXIMATELY HERE)

Figures A2 and A3 in the Appendix depict the *h*-index values obtained by the sensitivity analysis on the complete data. For the varying levels of P, C I utilize percentiles from the overall publication/citation distribution in each field (i.e. the 5%, 10%, 25%, 50%, 75%, 90% and 95%). It is clear from the inspection of the graphs that the *h*-index as a function of the *C*- and *P*-percentiles shows a convex trend. The corresponding progressive values of *SI*s for the two journal categories are shown in Figures A4 and A5. Interestingly, it is the selected distribution for the data that mostly characterizes the shape of the curves for the *h*-index and the *SI*s, rather than the mathematical functions.

The results in Tables 4 and 5 and the corresponding figures indicate that all models exhibit similar patterns in terms of sensitivity of the *h*-index to the varying levels of *P*, *C*. However, higher and lower levels of sensitivity are shown by the Glänzel-Schubert model. Indeed, less sensitive to changes is the Glänzel-Schubert model that measures changes in the number of publications (see Figures A2e – A5e), keeping the number of citations at their median. In addition, especially constraining the sensitivity analysis at a local range around the median values of *C*, *P* we find that the Glänzel-Schubert model is the most sensitive in terms of citations (see Table 5). A general behavior seems to be followed in all tested models where the *h*-index is approximately robust up to the 10-25$^{th}$ percentile of the total citation distribution of the journals in the specific field and then inflates (see Figures A2, A3). This is also verified by the inspection of the progress of *SI*s (Figures A4 and A5), where the increase of the *h*-index's sensitivity is apparent between the 10$^{th}$-25$^{th}$ percentile of the *C* distribution. A similar fluctuation is observed at the 95$^{th}$ percentile of the citation distribution in both datasets. Sensitivity with regard to changes in publications does not increase after the 95$^{th}$ percentile of the *P* distribution, as is the case with *C*. Again this characteristic applies to both datasets used for our analyses. The similar patterns of sensitivity in both fields of research could serve as a guide for the editor of any given journal on steps (s)he should undertake in order to grow the *h*-index of the journal. Hence, it seems that in order to substantially increase a journal's *h*-index, it is required the journal's citation output to reach at least the citations of the 10$^{th}$ percentile of the citation distribution in the specific field (e.g., approximately 52 and 755 citations for forestry and ecology, respectively) or, in other words, the journal should be among the 90% most cited journals. Similarly, retaining the number of citations at a constant level, one has to achieve a large reduction in publications to achieve a significant increase in *h*.

## 4 Conclusions

This paper aimed at investigating for the first time in a systematic way the sensitivity of the journal *h*-index to changes in the number of publications and citations, by utilizing some of the most popular *h*-index mathematical models. In doing this, the Bayesian paradigm was followed as regards the statistical methodology, offering a different view of model-based inference compared to classical statistics. The results showed that a model, including both *C* and *P* as explanatories for *h* is more sensitive to changes in *C* than models relying solely on *C*, especially when restricting these changes locally around the median *C*. Examination of the *h*-index's sensitivity to large variations of *C*, *P* revealed a pattern for almost all models, with the *h*-index values exhibiting robust behavior up to a certain value for *C* and inflating afterwards. Robustness of the *h*-index is considerably more apparent for *P* in comparison to *C*, verifying the theoretical beliefs. Clearly, the number of citations shows a stronger effect on the *h*-index in comparison to the number of publications, at least when utilizing the Glänzel-Schubert mathematical function.

Recognizing that although this initial analysis is far from exhaustive, the latter findings may prove useful for editorial policies, for instance, in assisting journal editors that seek to increase the reputation of their journal in terms of their respective *h*-index values. In doing this, the particular estimates from each field could be utilized for testing scenarios associated with varying the input parameters of *P*, *C* and gaining insight on how a specific journal's *h*-index varies according to these scenarios. The proposed methodology is currently applied also to other *h*-type indices for comparing their robustness with that of the *h*-index and this is a subject of ongoing research.

**TABLES**

| model | parameters | range | reference |
|---|---|---|---|
| $h = \left(\dfrac{\alpha-2}{\alpha-1} A\right)^{1/\alpha}$ | $\alpha$ | $\alpha \in (2, \infty)$ | Egghe-Rousseau (*Egghe and Rousseau, 2006*) |
| $h = cP^{1/(\alpha+1)} (C/P)^{\alpha/(\alpha+1)}$ | $(\alpha, c)$ | $\alpha \in (1, \infty); c \in (0, \infty)$ | Glänzel-Schubert model (*Schubert and Glänzel, 2007*) |
| $h = \left(\dfrac{C}{\alpha}\right)^{1/ab}$ | $(\alpha, a, b)$ | $\alpha, a \in (1, \infty); b = f(a)$ | Three-parameter Hirsch model *(Ye, 2011)* |

**Table 1.** *Theoretical models for the h-index based on P and C*

| Model | Distribution | $\bar{D}$ (ecology) | $\bar{D}$ (forestry) |
|---|---|---|---|
| Glänzel-Schubert (G-S) | Gaussian | 894.4 | 302.2 |
| Egghe and Rousseau (E-R) | NB | 1044 | 321.6 |
| Three-parameter Hirsch (H G) | Gaussian | 1020 | 328.5 |
| Three-parameter Hirsch (H NB) | NB | 960.1 | 310.7 |

**Table 2.** *Mean deviance ($\bar{D}$) for the fitted models*

| Model | Distribution | parameters | | | | | | | |
|---|---|---|---|---|---|---|---|---|---|
| | | (ecology) | | | | (forestry) | | | |
| | | *a* | a | *b* | *c* | *a* | a | *b* | *c* |
| Egghe-Rousseau | NB | 5.346 (5.25-5.44) | -- | -- | -- | 5.798 (5.58-6.03) | -- | -- | -- |
| Glänzel-Schubert | Gaussian | 1.77 (1.65-1.89) | -- | -- | 0.7 (0.64-0.75) | 1.966 (1.68-2.3) | -- | -- | 0.784 (0.65-0.95) |
| Three-parameter Hirsch | Gaussian | 4.362 (2.64-6.84) | 2.129 (0.76-5.03) | 1.01 (0.41-2.72) | -- | 4.499 (3.07-5.92) | 1.889 (1.02-4.45) | 1.265 (0.49-2.5) | -- |
| Three-parameter Hirsch | NB | 0.156 (0.08-0.29) | 1.372 (1.09-1.71) | 6.299 (5.06-7.9) | -- | 1.166 (0.47-2.93) | 1.103 (0.88-1.46) | 6.62 (5.09-7.93) | -- |

**Table 3.** *Posterior parameter estimates (medians) along with the 95% credible intervals of the theoretical models for the h-index, obtained from simulation of 50,000 values*

| Model | SI (ecology) | SI (forestry) |
|---|---|---|
| Egghe and Rousseau (NB) | 0.994 | 0.986 |
| Three-parameter Hirsch (Gaussian) | 0.986 | 0.988 |
| Three-parameter Hirsch (NB) | 0.988 | 0.980 |
| Glänzel-Schubert (Gaussian) C | 0.998 | 0.999 |
| Glänzel-Schubert (Gaussian) P | 0.765 | 0.867 |

**Table 4.** *Sensitivity index for the tested models (complete data output)*

| Model | SI (ecology) | SI (forestry) |
|---|---|---|
| Egghe and Rousseau (NB) | 0.301 | 0.236 |
| Three-parameter Hirsch (Gaussian) | 0.246 | 0.225 |
| Three-parameter Hirsch (NB) | 0.267 | 0.218 |
| Glänzel-Schubert (Gaussian) C | 0.327 | 0.337 |
| Glänzel-Schubert (Gaussian) P | 0.158 | 0.184 |

**Table 5.** *Sensitivity index for the tested models (-30% to 30% range of P, C values around their medians)*

**APPENDIX**

|        | h      | P      | C        |
|--------|--------|--------|----------|
| min    | 2      | 48     | 19       |
| 5%     | 7      | 145.4  | 291      |
| 10%    | 11.9   | 215    | 754.6    |
| 25%    | 25.25  | 565    | 3651.5   |
| median | 45.5   | 1351   | 14917.5  |
| 75%    | 83.75  | 2875.5 | 46843    |
| 90%    | 122    | 4631.8 | 143452.1 |
| 95%    | 153.3  | 6560.6 | 193911.8 |
| max    | 246    | 8678   | 456498   |

**Table A1.** *Descriptive statistics for the h-index, P and C (Ecology journals)*

|        | h     | P      | C        |
|--------|-------|--------|----------|
| min    | 1     | 18     | 3        |
| 5%     | 2     | 46.9   | 20.65    |
| 10%    | 3     | 69     | 51.6     |
| 25%    | 4.25  | 173.25 | 122.5    |
| median | 19    | 405.5  | 2435     |
| 75%    | 39    | 1616   | 13743.75 |
| 90%    | 73.7  | 3173.5 | 44343.2  |
| 95%    | 85.8  | 6116   | 63350.15 |
| max    | 101   | 8374   | 135245   |

**Table A2.** *Descriptive statistics for the h-index, P and C (Forestry journals)*

```
model{

#likelihood
for(i in 1:130){h[i] ~ dnorm(mu[i],tau.wth)I(0,)
mu[i]<-c*pow(P[i],b)*pow((C[i]/P[i]),d)}

b<-1/(a+1)
d<-a/(a+1)

# prior distributions
c~dnorm(0,0.01)I(0,)
a~dnorm(1,0.01)I(0,)
tau.wth ~ dgamma(0.001,0.001)
swth <- 1/tau.wth
}

#initial values
list(c=1,a=1, tau.wth=1)
```

**Fig A1.** *Example of WinBUGS code specifying the prior distributions and the likelihood for computing the posterior distribution of parameters $\theta = (\alpha, c)$ for the Glänzel-Schubert model.*

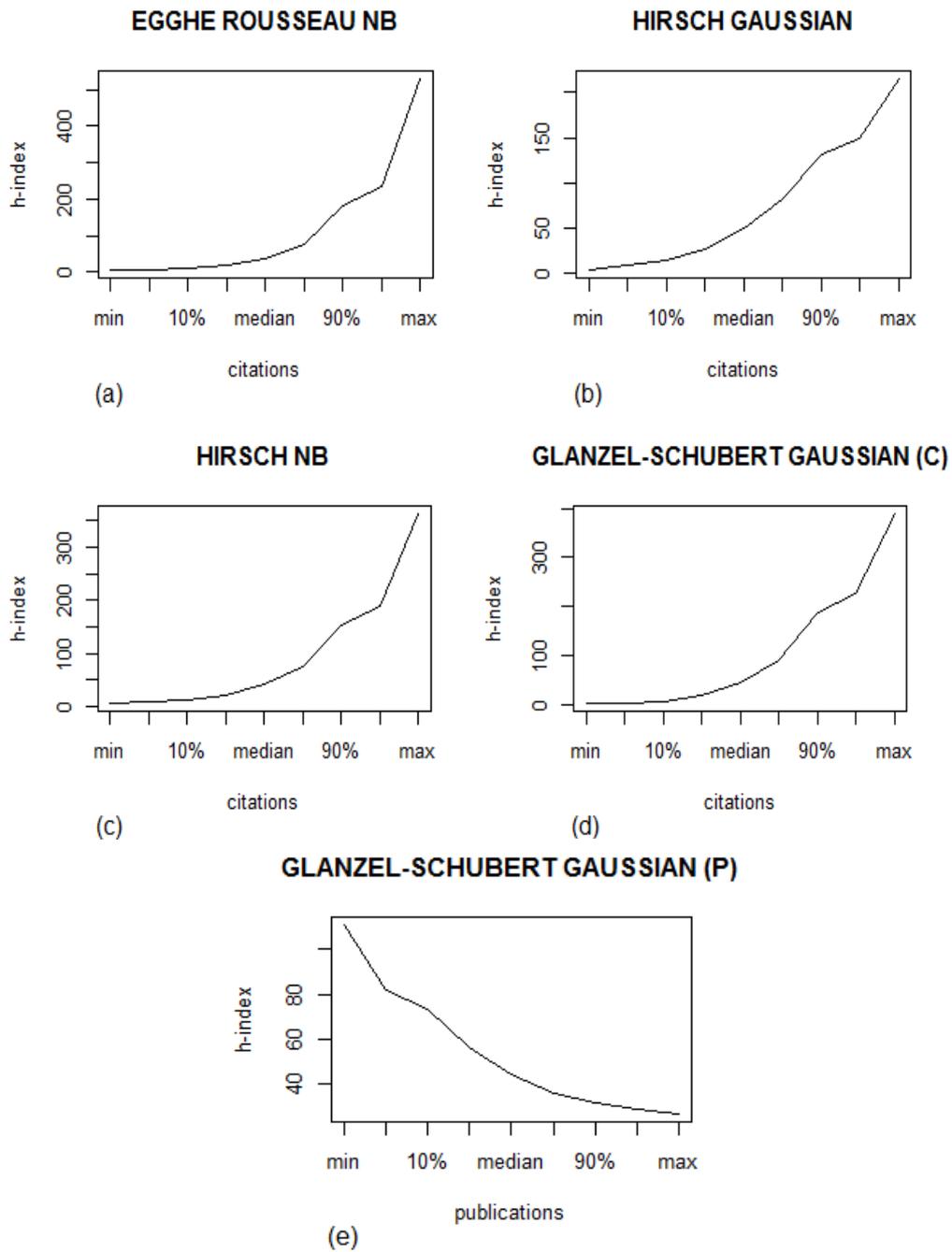

**Fig A2.** *Sensitivity analysis graphs for the h-index based on percent of publications/citations distribution (WoS Ecology journals)*

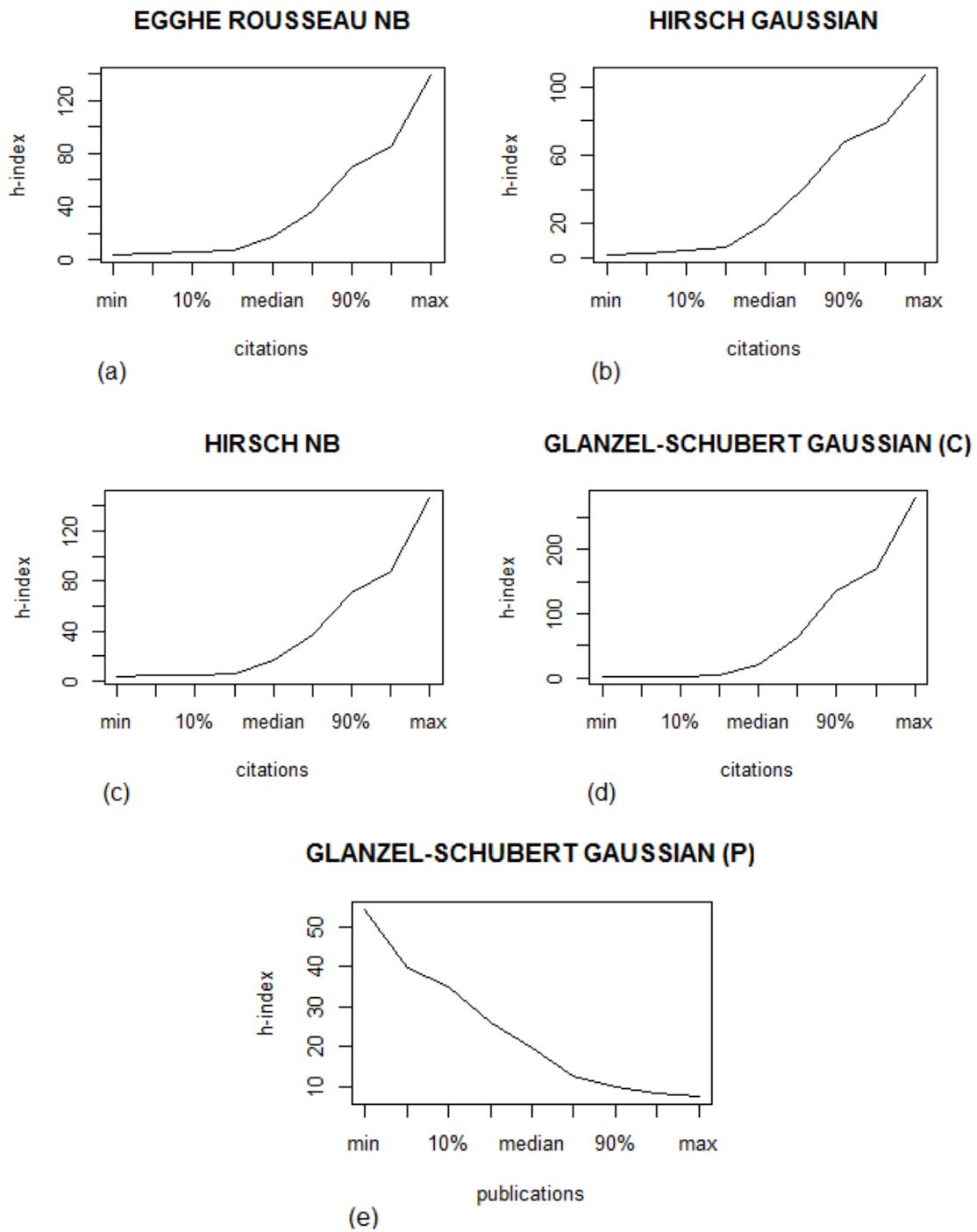

**Fig A3.** *Sensitivity analysis graphs for the h-index based on percent of publications/citations distribution (WoS Forestry journals)*

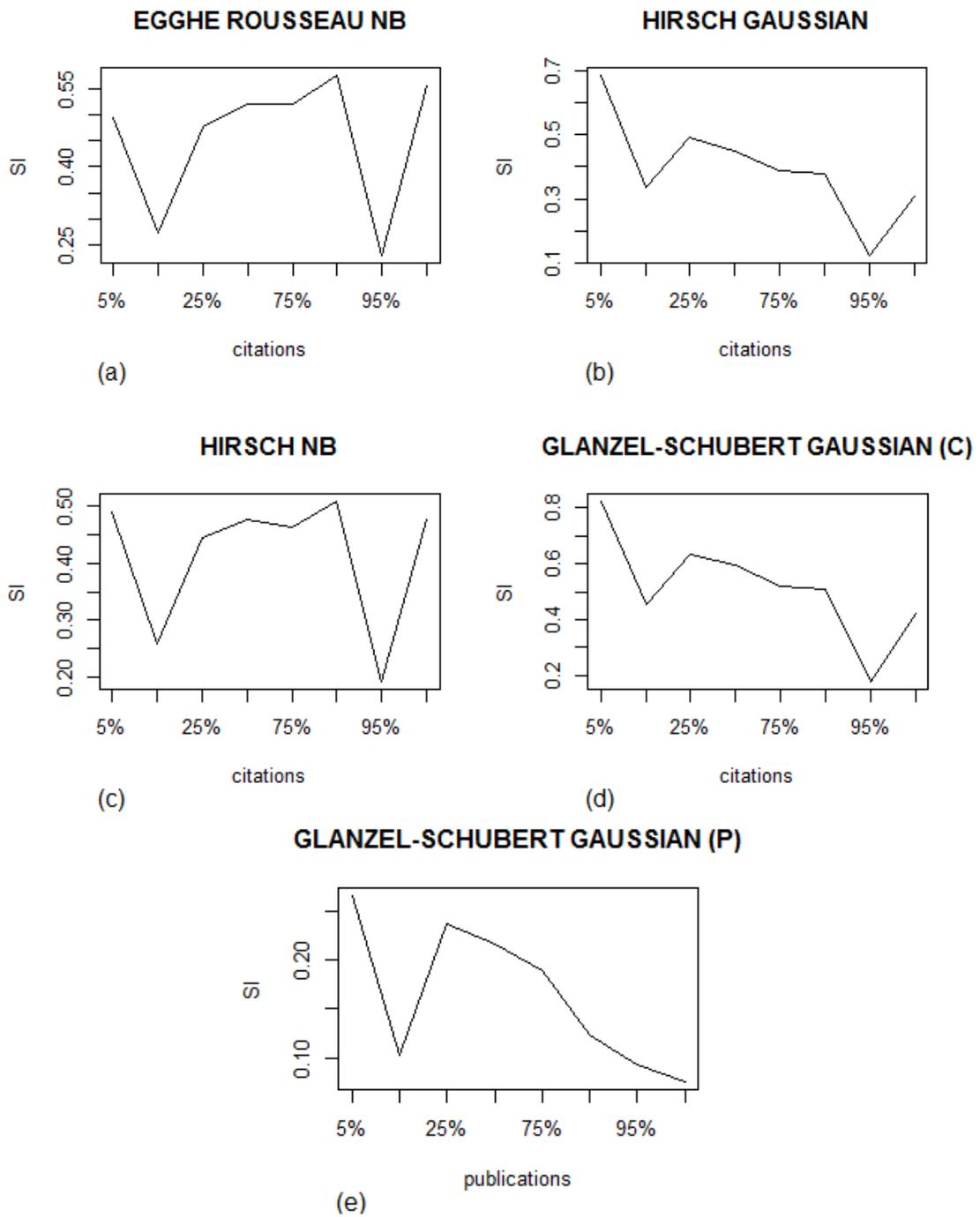

**Fig A4.** *SIs for the h-index based on percent of publications/citations distribution (WoS Ecology journals)*

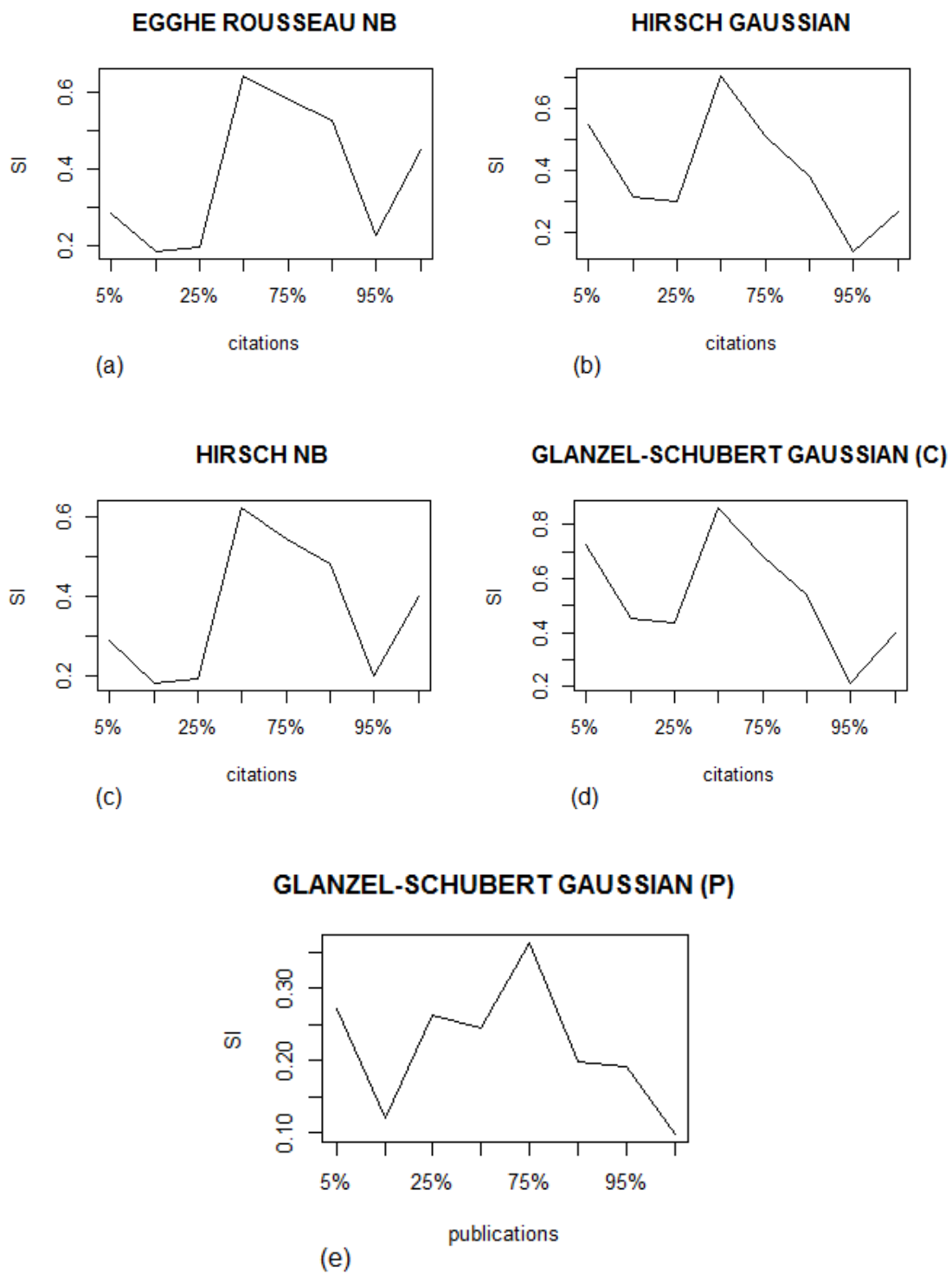

**Fig A5.** *SIs for the h-index based on percent of publications/citations distribution (WoS Forestry journals)*